\documentclass{article}
\usepackage{spconf}

\usepackage{color}
\usepackage{bm}
\usepackage{hyperref}

\usepackage{color}
\usepackage{theorem}
\usepackage{times,amsmath}
\usepackage{amssymb}
\usepackage{subcaption}
\usepackage{cite}
\usepackage{url}
\usepackage[capitalise]{cleveref}

\ninept

\usepackage[ruled,linesnumbered]{algorithm2e}

\let\forallsymb\forall

\input{mysymbol.sty}

\usepackage{tikz, pgfplots}
\usetikzlibrary{shapes,arrows}
\pgfplotsset{compat=1.17}
\pgfplotstableset{col sep=semicolon}
\usepackage{moresize} % provide more sizes for pgfplot
\usepgfplotslibrary{groupplots}
\tikzset{every mark/.append style={scale=1.6, solid}, font=\small}
\pgfplotsset{
    width=1\textwidth,
    height=5.5cm,
    legend style={
        font=\small ,  %\scriptsize,  %\ssmall,
        inner xsep=1pt,
        inner ysep=1pt,
        nodes={inner sep=1pt}},
    legend cell align=left,
    every axis/.append style={line width=.5pt},
 	every axis plot/.append style={line width=2.25pt},
 	every axis y label/.append style={yshift=-4pt}
}

\DeclareMathOperator{\Tr}{Trace}

\newtheorem{problem}{\bf Problem}

\title{Blind Deconvolution of Sparse Graph Signals in the Presence of Perturbations}
\name{Victor M. Tenorio, Samuel Rey, and Antonio G. Marques
\thanks{Work partially supported by the Spanish AEI (AEI/10.13039/ 501100011033) grants PID2019-105032GB-I00, PID2022-136887NB-I00 and FPU20/05554, and by the Young Researchers R\&D Project, ref. num. F861 (CAM and URJC). All the authors are with the Dept. of Signal Theory and Comms., King Juan Carlos University, Madrid, Spain. Email contact author: antonio.garcia.marques@urjc.es.}}
\address{Dept. of Signal Theory and Communications, King Juan Carlos University, Madrid, Spain}

\begin{document}
%\ninept
%
\maketitle
\begin{abstract}
Blind deconvolution over graphs involves using (observed) output graph signals to obtain both the inputs (sources) as well as the filter that drives (models) the graph diffusion process.
This is an ill-posed problem that requires additional assumptions, such as the sources being sparse, to be solvable.
This paper addresses the blind deconvolution problem in the presence of imperfect graph information, where the observed graph is a \emph{perturbed} version of the (unknown) true graph.
While not having perfect knowledge of the graph is arguably more the norm than the exception, the body of literature on this topic is relatively small.
This is partly due to the fact that translating the uncertainty about the graph topology to standard graph signal processing tools (e.g. eigenvectors or polynomials of the graph) is a challenging endeavor. To address this limitation, we propose an optimization-based estimator that solves the blind identification in the vertex domain, aims at estimating the inverse of the generating filter, and accounts explicitly for additive graph perturbations.
Preliminary numerical experiments showcase the effectiveness and potential of the proposed algorithm.
\end{abstract}

\section{Introduction}\label{S:Introduction}

In recent years, we have witnessed an exponential surge in data utilization, accompanied by a concurrent rise in its complexity.
Addressing this growing complexity involves harnessing the inherent structure within the data. 
In this context, Graph Signal Processing (GSP)~\cite{shuman2013emerging,djuric2018cooperative,ortega2018graph} emerges as a solution to this challenge, employing graphs to capture the underlying structure of the data and interpreting the data as signals defined on the nodes of the graph. 
%employing graphs to represent such structures and interpreting data as signals defined on the nodes of the graph.
This convergence of signal processing and graph theory facilitates navigating the intricacies of modern information, revealing insights and patterns that traditional methods might overlook.

Leveraging the structure of the data becomes even more relevant in ill-posed problems where the available observations are insufficient to solve the task at hand.
This is precisely the case of blind deconvolution of graph signals, which is an extension to graphs of the classical problem of blind system identification or blind deconvolution of signals in the time or spatial domain~\cite{pinto2012locating,ahmed14blindDeconv}. 
Specifically in GSP, given a set of output signals that are assumed to be the output of a diffusion process driven by a graph filter (GFi), blind deconvolution tries to jointly identify the sources (nodes) of the diffusion process as well as the GFi that drove the diffusion~\cite{pena2016source,segarra17blind,ramirez21blinddeconv,ye18blind,ye22learning}.
This problem finds applications in multiple domains, such as in social analysis (identifying the sources and the propagation dynamics of a rumor in a social network) or in neuroscience (identifying the sources of neurological activity as well as the diffusion patterns in a brain network)~\cite{wu2013blind}. %~\cite{kramer2008emergent,wu2013blind,feizi2019network}.

The aforementioned approaches to the blind deconvolution problem assume perfect knowledge of the graph topology. This simplifying assumption allows to compute the frequency representation of the graph signals and the diffusing GFi, leveraging those representations in their proposed algorithms.
Nonetheless, in real-world applications, the graph is prone to contain \emph{imperfections} due to, for example, noise in the observed links or outdated information when the graph varies with time.
%~\cite{egeland2009availability}.
Furthermore, when in lieu of physical entities the graph captures pairwise relationships that need to be learned from the data, the limitations of the method used to learn the topology may give rise to perturbations.
%~\cite{friedman2008sparse,dong2016learning}.
The ubiquity of graph perturbations and their potential impact on the performance of graph-based algorithms highlight the necessity to develop robust algorithms capable of effectively handling imperfect graph information. Unfortunately, this is not a trivial task, since even characterizing the influence that simple perturbations models (e.g., additive noise models) have on classical GSP tools (the eigenvectors of the graph-shift operator (GSO) that define the graph Fourier transform or the powers of the GSO that are used in a GFi) is quite challenging.
%As a consequence, there has been a clear increase in the development of robust approaches across various domains within GSP, such as the fields of filter identification~\cite{rey21icassprgfi,rey23rgfi} or graph neural networks~\cite{tenorio2021robust,jin2020graph}.
Despite these challenges, several GSP works have started to look into this relevant limitation.
Works in~\cite{ceci2020graph,nguyen2022stability} study the influence of perturbations in the spectrum of the graph while~\cite{miettinen2018graph} focus on the effect of perturbations in GFis of order one.
Rather than analyzing the effects of perturbations,~\cite{ceci2020_semtls,natali2020topology,rey23rgfi} address the identification of (different types of) GFis while accounting for imperfections in the observed topology.
Finally, the presence of perturbations has also been considered in non-linear methods, where current approaches range from studying the transferability of non-linear GFis to designing novel GFis robust to perturbations~\cite{jin2020graph,tenorio2021robust,levie2021transferability}.
However, notwithstanding the rising focus on the uncertainty in the graph topology, there have been no efforts to tackle the problem of blind deconvolution from a robust standpoint.

\noindent \textbf{Contributions.} 
To address the previous limitations, this paper poses the task of robust blind deconvolution of graph signals while considering imperfect topology knowledge.
It carefully formulates the problem as a non-convex optimization program and, by relying on different convex relaxations, designs an algorithm to find a solution.
The key aspects of the proposed approach are: 1)~modeling the true (unknown) graph as an explicit optimization variable to account for perturbations in the observed topology; 2)~optimizing over the inverse filter rather than the generating one to simplify the objective function; and 3)~modeling the dependence of the inverse GFi on the graph via a commutativity constraint in the vertex domain, which bypasses the challenges of dealing with high-order polynomials and working on spectral domain of the graph. The postulated problem uses the observations of the output signals and an imperfect version of the GSO to jointly obtain the sources of the network diffusion, the underlying GFi driving the process, and an enhanced (denoised) estimate of the graph. Since the sources of non-convexity are reduced to a mere bilinearity, the optimization is tackled through an alternating approach, labeled as the Robust Blind Deconvolution over Graphs (RBDG) algorithm.
To the best of our knowledge, this is the first work that does not assume perfect knowledge of the graph topology when solving the blind identification problem.

\section{Blind Deconvolution in GSP}\label{S:notation-relatedwork}

In this section, we introduce notation and some fundamentals of GSP. Then, we discuss blind deconvolution of graph signals, and present how previous works dealt with it.

\vspace{1.5mm}
\noindent \textbf{Notation and GSP preliminaries.} Denote by $\ccalG = (\ccalV, \ccalE)$ a graph with $\ccalV$ and $\ccalE$ representing its node and edge set, respectively, and with $|\ccalV| = N$ nodes. Denote by $\bbA \in \reals^{N \times N}$ its (possibly weighted and directed) adjacency matrix, where $A_{ij} \neq 0$ if and only if $(i,j) \in \ccalE$. More generally, GSO $\bbS \in \reals^{N \times N}$ is a matrix encoding the structure of the graph, where $S_{ij}$ can be non-zero only if $(i,j) \in \ccalE$ or if $i=j$. Classical examples of matrices playing the role of the GSO are the adjacency matrix or the combinatorial graph Laplacian $\bbL = \text{diag} (\bbd) - \bbA$, where the entries $d_j = \sum_j A_{ij}$ represent the nodal degrees.
Define also a graph signal as the mapping $\ccalV \to \reals$, which can be conveniently represented by a vector $\bbx \in \reals^N$, where the entry $x_i$ encodes the signal value at node $i \in \ccalV$.
Finally, a fundamental role in graph signal deconvolution is played by GFis.
A GFi is a graph-aware linear operator for graph-signals that can be represented as a matrix polynomial of the GSO of the form 
\begin{equation}\label{eq:graph_filter}
    \bbH = \sum_{r=0}^{N-1} h_r \bbS^r,
\end{equation}
with $\bbh = [h_0, ..., h_{N-1}]$ denoting the vector of filter coefficients \cite{sandryhaila2014discrete}.
Since $\bbS^r$ encodes the r-hop neighborhood of the graph, GFis are widely used to model diffusion processes over networks~\cite{segarra17blind}.

\vspace{1.5mm}
\noindent \textbf{Blind deconvolution of graph signals.}
Consider a diffusion process where the output signal $\bby \in \reals^{N}$ is given by
%Consider a graph signal $\bby \in \reals^{N \times N}$, obtained from the diffusion of a sparse signal $\bbx$ across the graph according to
\begin{equation}\label{eq:diff_model}
    \bby = \bbH\bbx + \bbw,
\end{equation}
with $\bbx \in \reals^{N}$ being the input signal being diffused, $\bbH$ a GFi modeling the diffusion process, and $\bbw \in \reals^{N}$ a random vector representing noise or model inaccuracies.
Then, given a set of $M$ observed signals $\{\bby_i\}_{i=1}^M$ generated according to \eqref{eq:diff_model}, blind deconvolution aims to find the sources of the network diffusion $\{\bbx_i\}_{i=1}^M$ as well as the GFi $\bbH$ controlling the diffusion process.
This is a challenging and ill-posed problem since both $\bbx$ and $\bbH$ are unknown.
Therefore, to promote its tractability, a workhorse approach is to assume that there are only a few sources for the network diffusion  (i.e. $\bbx_i$ are sparse).
Moreover, exploiting that $\bbH$ is a GFi so only the coefficients $\bbh$ are unknowns becomes critical.

Early works dealing with the blind deconvolution problem in the context of GSP appear in~\cite{segarra17blind,yuejie2016blinddeconvInversefilter}. The approach put forth recovers a lifted rank-one matrix $\bbZ = \bbx \bbh^T$, which exhibits certain desirable properties such as being row sparse and rank one. Later on, the works presented in~\cite{iglesias2018demixing,ramirez21blinddeconv} review several existing methods and extend the previous approach to input signals defined as the combination of a few entries in a dictionary, as well as exploring the problem of graph signal sampling and analyzing its similarities with blind deconvolution.

Differently,~\cite{yuejie2016blinddeconvInversefilter,ye18blind} reformulate the problem to identify the frequency response of the inverse filter as a way to bypass the non-convex bilinear term $\bbH \bbx_i$ that arises when jointly identifying the filter and the input signals.
The work in~\cite{ye18blind} is further developed in~\cite{ye22learning}, where the authors incorporate unrolling schemes~\cite{monga21unrolling} to strengthen their design and limit the impact of selecting the hyperparameters.

%\section{Problem and Methodology}
\section{Robust blind deconvolution over graphs}\label{S:methodology}

%In this section we will formally state the problem we are trying to solve, and we will also propose an algorithm to solve the mentioned problem.
After having established the notation and the problem context, along with an overview of prior approaches to address it, this section presents the formal problem statement and introduces our proposed algorithmic solution.

As previously mentioned, we assume that we do not have access to the true GSO, but to a noisy version $\barbS = \bbS + \bbDelta$.
Here, $\bbDelta$ represents a perturbation matrix whose particular structure will depend on the perturbation at hand (e.g., creating/destroying links, or noisy edge weights)~\cite{rey23rgfi}.
It is easy to note that the uncertainty encoded in $\bbDelta$ renders the blind deconvolution problem more challenging to solve.
The blind identification problem accounting for the graph imperfections is formally stated next.

\begin{problem}
Let $\ccalG$ be a graph with $N$ nodes, $\bbS$ the true (unknown) GSO, and $\barbS$ be the perturbed (observed) GSO.
Moreover, let $\{\bby_i\}_{i=1}^M$ be the observed output signals obtained from the unknown input signals $\{\bbx_i\}_{i=1}^M$ as described in \eqref{eq:diff_model}.
The aim is to use the duplet $(\{ \bby_i\}_{i=1}^M, \barbS )$ to find i) the input signals $\{\bbx_i\}_{i=1}^M$; ii) the filter driving the diffusion process $\bbH$; and iii) an enhanced estimate of the real GSO $\bbS$. To that end, the following assumptions are in order:

\noindent(\textbf{AS1}) The input signals $\bbx_i$ are sparse, i.e., $\| \bbx_i \|_0 = K \; \forallsymb \; i \in \{1, ..., M\}$, where $K \ll N$.

\noindent(\textbf{AS2}) $\bbH$ is a polynomial of $\bbS$.

\noindent(\textbf{AS3}) $\bbS$ and $\barbS$ are close according to some metric $d(\bbS,\barbS)$, i.e., the observed perturbations are ``small'' in some sense.

\label{prob1}
\end{problem}

Assumptions (\textbf{AS1}) and (\textbf{AS2}), which limit the degrees of freedom of the bilinear model, are standard in the context of blind deconvolution of graph signals~\cite{segarra17blind,ye18blind}. By limiting the level of perturbations, Assumption (\textbf{AS3}) guarantees that matrices $\bbS$ and $\barbS$ are  ``similar'' and, as a result, that $\barbS$ contains meaningful information about $\bbS$. 

For convenience, let us group the input and output signals in the columns of the $N \times M$ matrices $\bbX := [\bbx_1,..., \bbx_M]$ and $\bbY := [\bby_1, ..., \bby_M]$, respectively. A natural optimization-based formulation for Problem~\ref{prob1} is

\begin{align}
    \min_{\mathbf{X, H, S}} & \| \bbY - \bbH \bbX \|_F^2 + \beta \| \bbS \|_0 \label{eq:orig-obj}\\
    \text{s.to:} \; \; & d(\bbS, \barbS) \leq \epsilon_1 \label{eq:orig-const1} \\
    & \|\bbx_i \|_0 \leq K \; \forallsymb \; i \in \{1,...,M\} \label{eq:orig-const2} \\
    & \|\mathbf{HS - SH} \|_F^2 \leq \epsilon_2, \label{eq:orig-const3}
\end{align}
where we formulated the objective function to minimize the error between the output signals $\bbY$ and the prediction $\bbH \bbX$, along with an $\ell_0$ term of the GSO to promote a sparse solution for $\bbS$. The key constraint in our approach is~\eqref{eq:orig-const1}, which relates to (\textbf{AS3}) and bounds the distance between the GSO $\bbS$ and the observation $\barbS$.
The choice of a suitable distance function will depend on the nature of the perturbation encoded in $\bbDelta$, with plausible examples being the $\ell_0$ pseudo-norm when perturbations create/destroy links or the Frobenius norm when the weights of $\barbS$ present noise.
Next, the constraint~\eqref{eq:orig-const2} is used to limit the sparsity of the signals $\bbX$ while \eqref{eq:orig-const3} promotes that $\bbH$ is a GFi (i.e., a polynomial on $\bbS$) as stated in (\textbf{AS2}).
Note that the commutativity exploits the fact that, since $\bbH$ is a polynomial of $\bbS$, the two matrices share the same eigenvectors.
This simple observation prevents us from dealing with the spectrum of $\bbS$ and with high-order polynomials, simplifying the optimization problem.

Nevertheless, \eqref{eq:orig-obj} is clearly a non-convex optimization problem, due to the bilinear terms in~\eqref{eq:orig-obj} and \eqref{eq:orig-const3} and the $\ell_0$ norms in \eqref{eq:orig-obj} and \eqref{eq:orig-const2}. To design a convex solution, we rely on an alternating optimization approach.
To that end, we make the following considerations:
\begin{itemize}
    \item Inspired by~\cite{ye18blind,ye22learning}, we replace the optimization variable $\bbH$ by its inverse, denoted by $\bbG := \bbH^{-1}$. This change of variable simplifies the objective function by replacing the bilinearity $\|\bbY - \bbH\bbX\|_F^2$ with $\| \bbG \bbY - \bbX \|_F^2$, which is convex in both $\bbG$ and $\bbX$.
    %aims to remove the bilinear term from the objective function (first term in~\eqref{eq:orig-obj}), so that we can rewrite it as $\| \bbG \bbY - \bbX \|_F^2$, which is convex in both $\bbG$ and $\bbX$.
    A key aspect to note is that this variable change still allows us to use the commutativity term, since the inverse filter $\bbG$ also shares eigenvectors with $\bbS$.\footnote{To demonstrate this, let us write the generating GFi as $\bbH = \bbV \text{diag} (\tbh) \bbV^{-1}$, where $\bbV$ are the eigenvectors of $\bbS$ and $\tbh$ is the frequency response of the filter. We can then write $\bbG \!= \!\bbV \text{diag} (\tbg) \bbV^{-1}$, where $\tilde{g}_i \!= \! 1 / \tilde{h}_i$ for all $i$. Therefore, $\bbG$ and $\bbS$ share eigenvectors, and thus commute.}
    \item We replace the $\ell_0$ norm by its convex surrogate, the $\ell_1$ norm (to simplify exposition, iterative re-weighted $\ell_1$ alternatives~\cite{candes2008enhancing} are not included in the formulation but they are considered in the numerical section).
    \item We move the constraints to the objective function by adding them as regularizers.
    \item We assume that $\barbS$ contains perturbations that create and destroy links of $\bbS$ and select the $\ell_1$ norm as the distance function between $\barbS$ and $\bbS$.
    Nevertheless, recall that any other convex distance can be readily employed.
\end{itemize} 

Taking into account the previous considerations, we end up with the following optimization problem
\begin{align}
\min_{\bbX, \bbG, \bbS} & \| \bbG \bbY - \bbX \|_F^2 + \beta \| \bbS \|_1 + \lambda \| \bbS - \barbS\|_1 \label{eq:prob-reform} \\
& + \alpha \| \bbX \|_1 + \gamma\|\bbG \bbS - \bbS \bbG \|_F^2 \nonumber \\
\text{s.to:} \;\; & \Tr(\bbG) = 1, \nonumber
\end{align}
where the constraint is used to prevent the trivial solution ($\bbX=0,\bbG=0,\bbS=\barbS$). The problem in \eqref{eq:prob-reform} is still non-convex due to the bilinearity in the commutativity term, but can be solved iteratively using an alternating minimization approach where, at each iteration $t$, two steps are performed:

\begin{itemize}
    \item \textbf{Step 1} (filter and source identification): in this step, we find the optimal solutions for both $\bbG$ and $\bbX$ by solving the following problem

    \begin{align}
        \bbG_{(t)}, \bbX_{(t)} = \text{arg} \min_{\bbG, \bbX} \; & \| \bbG \bbY - \bbX \|_F^2 + \alpha \|\bbX \|_1 \label{eq:step1} \\
        & + \gamma\|\bbG \bbS_{(t-1)} - \bbS_{(t-1)} \bbG \|_F^2 \nonumber \\
        \text{s.to:} \;\; & \Tr(\bbG) = 1, \nonumber
    \end{align}
    where we used the estimation of $\bbS$ from the previous iteration, $\bbS_{(t-1)}$. This problem, which is convex in both $\bbG$ and $\bbX$, can be solved using standard convex solvers. However, more efficient approaches, such as coordinate descent or proximal methods can be employed~\cite{hastie2015statistical}.
    
    \item \textbf{Step 2} (graph denoising): with the solutions of the previous step, we now aim to find a new estimation of the GSO by solving

    \begin{align}
    \bbS_{(t)} = \text{arg} \min_{\bbS} \; & \beta \| \bbS \|_1 + \lambda \| \bbS - \barbS\|_1 \label{eq:step2} \\
    & + \gamma\|\bbG_{(t)} \bbS - \bbS \bbG_{(t)} \|_F^2, \nonumber
    \end{align}
    which is also convex and amenable to efficient approaches like the previous step.
\end{itemize}

The complete pseudo-code is presented in Algorithm~\ref{A:rbd_alg}. The two steps~\eqref{eq:step1}-\eqref{eq:step2}, which are coupled by the commutativity term, are repeated for a fixed number of iterations $T$ or until a stopping criterion is met. For the first iteration, the GSO is initialized to the imperfect observation $\bbS_{(0)} = \barbS$. The output of Algorithm~\ref{A:rbd_alg} are the estimates for the inverse filter $\hbG = \bbG_{(T)}$, the source signals $\hbX = \bbX_{(T)}$, and the denoised GSO $\hbS = \bbS_{(T)}$.

\begin{algorithm}[tb]
\SetKwInput{Input}{Input}
\SetKwInOut{Output}{Output}
\Input{$\bbY$, $\barbS$}
\Output{$\hbG$, $\hbX$, $\hbS$.}
\SetAlgoLined
Initialize $\bbS_{(0)}$ as $\bbS_{(0)} = \barbS$. \\
\For{$t=1$ \KwTo $T$}{
    Compute $\bbG_{(t)}$ and $\bbX_{(t)}$ by solving \eqref{eq:step1} using $\bbS_{(t-1)}$. \\
    Compute $\bbS_{(t)}$ by solving \eqref{eq:step2} using $\bbG_{(t)}$ and $\bbX_{(t)}$.
    \\
}
$\hbG = \bbG_{(T)},\; \hbX = \bbX_{(T)},\; \hbS = \bbS_{(T)}$.
\caption{Robust blind deconvolution with graph denoising.}
\label{A:rbd_alg}
%\vspace{-.2cm}
\end{algorithm}

Note that, unlike~\cite{ye18blind}, we solve the problem in the vertex domain, without relying on the frequency representation of the signals. This allows us to bypass the instability that the perturbations in $\barbS$ generate in the graph eigenvectors, yielding a more robust approach. Another advantage of the proposed algorithm is that, unlike~\cite{segarra17blind}, we do not require the sparsity pattern of the sources to be the same across all signals.

Algorithm~\ref{A:rbd_alg} incorporates four essential hyperparameters: $\alpha$, $\gamma$, $\beta$, and $\lambda$. These hyperparameters can be found via grid search or any other hyperparameter search algorithm. Since $(\alpha, \gamma, \beta, \lambda)$ are associated with regularizers derived from constraints, they can be interpreted as Lagrange multipliers; as a result, their values can be potentially tuned using tailored dual algorithms. Alternatively, an approach based on the unrolling technique, like the one proposed in~\cite{ye22learning}, could be implemented. Several of these options will be explored and compared in the journal version of this paper.

\section{Numerical experiments}\label{S:exps}

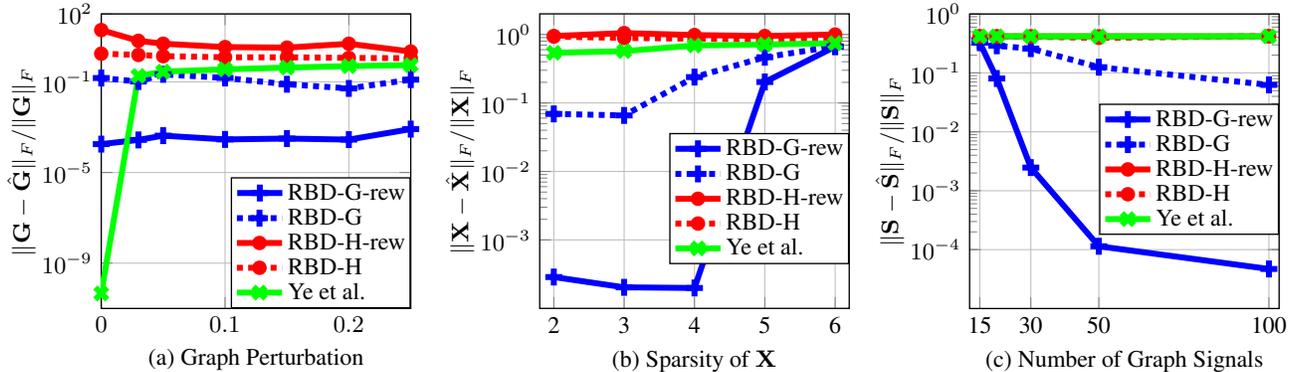
\begin{figure*}[!t]
	\centering
	\begin{subfigure}[t]{0.32\textwidth}
		\centering
		\begin{tikzpicture}[baseline,scale=1]

\begin{semilogyaxis}[
    table/col sep=comma,
    xlabel={(a) Graph Perturbation},
    xmin={0},
    xmax={0.25},
    ylabel={$\| \bbG - \hbG \|_F / \| \bbG \|_F$},
    ymin={1e-11},
    ymax={100},
    ytick={1e-9, 1e-5, 1e-1, 1e2},
    grid=major,
    legend style={
        at={(1,0)},
        anchor=south east},
    legend columns=1,
    ]
    
    \pgfplotstableread{data/err_G_pert.csv}\errGPert
    
    \addplot[blue, mark=+] table [x=Eps, y=RBD-G-rew] {\errGPert};
    \addplot[blue, dotted, mark=+] table [x=Eps, y=RBD-G] {\errGPert};
    \addplot[red, mark=*, mark size=1pt] table [x=Eps, y=RBD-H-rew] {\errGPert};
    \addplot[red, dotted, mark=*, mark size=1pt] table [x=Eps, y=RBD-H] {\errGPert};
    \addplot[green, mark=x] table [x=Eps, y={Ye et al.}] {\errGPert};
    
    \legend{RBD-G-rew,RBD-G,RBD-H-rew,RBD-H,{Ye et al.}}
    
\end{semilogyaxis}
\end{tikzpicture}
	\end{subfigure}
	\begin{subfigure}[t]{0.32\textwidth}
		\centering
		\begin{tikzpicture}[baseline,scale=1]

\begin{semilogyaxis}[
    table/col sep=comma,
    xlabel={(b) Sparsity of $\bbX$},
    xmin={1.8},
    xmax={6.2},
    xtick={2, 3, 4, 5, 6},
    xticklabels={2, 3, 4, 5, 6},
    ylabel={$\| \bbX - \hbX \|_F / \| \bbX \|_F$},
    ymin={1e-4},
    ymax={2},
    ytick={1e-3, 1e-2, 1e-1, 1e0},
    grid=major,
    legend style={
        at={(1,0.15)},
        anchor=south east},
    legend columns=1,
    ]
    
    \pgfplotstableread{data/err_X_sparsity.csv}\errGPert
    
    \addplot[blue, mark=+] table [x=S, y=RBD-G-rew] {\errGPert};
    \addplot[blue, dotted, mark=+] table [x=S, y=RBD-G] {\errGPert};
    \addplot[red, mark=*, mark size=1pt] table [x=S, y=RBD-H-rew] {\errGPert};
    \addplot[red, dotted, mark=*, mark size=1pt] table [x=S, y=RBD-H] {\errGPert};
    \addplot[green, mark=x] table [x=S, y={Ye et al.}] {\errGPert};
    
    \legend{RBD-G-rew,RBD-G,RBD-H-rew,RBD-H,{Ye et al.}}
    
\end{semilogyaxis}
\end{tikzpicture}
	\end{subfigure}
	\begin{subfigure}[t]{0.32\textwidth}
		\centering
		\begin{tikzpicture}[baseline,scale=1]

\begin{semilogyaxis}[
    table/col sep=comma,
    xlabel={(c) Number of Graph Signals},
    xmin={12},
    xmax={103},
    xtick={15, 30, 50, 100},
    xticklabels={15, 30, 50, 100},
    ylabel={$\| \bbS - \hbS \|_F / \| \bbS \|_F$},
    ymin={1e-5},
    ymax={1},
    ytick={1e-4, 1e-3, 1e-2, 1e-1, 1e0},
    grid=major,
    legend style={
        at={(1,0.25)},
        anchor=south east},
    legend columns=1,
    ]
    
    \pgfplotstableread{data/err_S_samp.csv}\errGPert
    
    \addplot[blue, mark=+] table [x=M, y=RBD-G-rew] {\errGPert};
    \addplot[blue, dotted, mark=+] table [x=M, y=RBD-G] {\errGPert};
    \addplot[red, mark=*, mark size=1pt] table [x=M, y=RBD-H-rew] {\errGPert};
    \addplot[red, dotted, mark=*, mark size=1pt] table [x=M, y=RBD-H] {\errGPert};
    \addplot[green, mark=x] table [x=M, y={Ye et al.}] {\errGPert};
    
    \legend{RBD-G-rew,RBD-G,RBD-H-rew,RBD-H,{Ye et al.}}
    
\end{semilogyaxis}
\end{tikzpicture}
	\end{subfigure}
		%\vspace{-0.15cm}
	\caption{Comparing the estimation performance of 5 blind deconvolution approaches.
	(a) shows the error in the recovered filter $\hbG$ when modifying the number of links perturbed in $\barbS$, where the values in the x-axis represent the proportion of the total number of links that have been perturbed,
	(b) represents the error in the sources of the diffusion $\bbX$ with respect to the increase in the number of active sources $K$, and
    (c) plots the error in the denoised GSO when increasing the number of samples $M$.}
    %Note that the approach in ``Ye et al.'' does not perform denoising, and therefore $\hbS = \barbS$ for this algorithm.}	%\vspace{-0.5cm}
    \label{F:exps}
\end{figure*}

In this section, we numerically assess the capabilities of the proposed algorithm. To do so, we first explain the setup common to all experiments and then we present and analyze the results for three test cases. The code used to run the simulations is available in GitHub\footnote{\url{https://github.com/vmtenorio/RobustBlindDeconvolution}}.

\noindent \textbf{Experiment setup}: Unless stated otherwise, we sample graphs with $N=20$ nodes from the small-world random graph model~\cite{newman1999renormalization}, we assign $K=2$ sources of the network diffusion by selecting $K$ nodes uniformly at random for each graph signal and we generate $M=50$ graph signals. We use the adjacency matrix as the GSO, and the coefficients of the GFi are sampled uniformly at random between 0 and 1. The graph is perturbed by rewiring a 10\% of the total number of edges. The error represented in the figures is the normalized median error across 25 realizations.

We compare Algorithm~\ref{A:rbd_alg} (labelled as ``RBD-G'' in the figure legends)
with: i) the scheme presented in~\cite{ye18blind} (``Ye et. al.'');
%where the frequency representation of the inverse filter obtained is used along with the eigenvectors of the perturbed GSO to obtain the matrix $\bbG$,
and ii) a slight modification of~\eqref{eq:orig-obj}-\eqref{eq:orig-const3}, where we augment the objective in \eqref{eq:orig-obj} with the constraints \eqref{eq:orig-const1}-\eqref{eq:orig-const3}, replace the $\ell_0$ pseudo norm with the $\ell_1$ norm, and set the distance function as the $\ell_1$ norm of the difference. The modified problem in ii) is solved using an iterative 3-step process where in step 1 we estimate the GFi $\bbH$, in step 2 we optimize $\bbX$, and in step 3 we estimate $\bbS$. This approach is labeled as ``RBD-H'' in the figures.
Finally, the lines whose labels contain ``-rew'' represent versions of the previous algorithms where the $\ell_1$ norms of $\bbX$ and $\bbS$ are replaced with a reweighted $\ell_1$-approach~\cite{candes2008enhancing}.

\noindent \textbf{Test case 1}:
Here, we analyze the impact of perturbations on the recovered GFi $\hbG$. Figure~\ref{F:exps}-(a) shows the normalized error as we increase the perturbations in $\barbS$. In general, we observe that the error grows as the ratio of rewired links increases and that the proposed approach outperforms the alternatives in every perturbed scenario. As we can see, for the unperturbed case (left-most point of the horizontal axis), the algorithm that yields the most accurate filter estimate is the approach in~\cite{ye18blind}, as we would expect. However, for the smallest perturbation value considered (second left-most point of the horizontal axis), the performance of the approach in~\cite{ye18blind} drops dramatically. In contrast, our robust algorithms, both using the reweighted version of the $\ell_1$ norm and without it, obtain results in the order of $10^{-4}$ and $10^{-1}$, respectively, for all perturbation values. In other words, they are able to properly deal with perturbations in the graph and consistently obtain an accurate representation of the GFi driving the process. Finally, it is worth noting that the naive 3-step approach is not able to properly identify the GFi, most probably due to the increased complexity of introducing a third step along with the non-convexity introduced by the additional bilinear term $\bbH \bbX$.

\noindent \textbf{Test case 2}:
Here our focus is on analyzing how the number of sources affects the quality of the estimation $\hbX$. 
Figure~\ref{F:exps}-(b) depicts the error in the recovered $\hbX$ as the number of non-zero entries ($K$) in each $\bbx_i$ increases.
From the results, it follows that our algorithms clearly outperform the alternatives.
More precisely, we observe that the error remains below $10^{-3}$ when $K < 5$, (which corresponds to 25\% of the total number of nodes in the graph), and then starts deteriorating.
Interestingly, comparing ``RBD-G-rew'' and ``RBD-G'', we observe that the reweighted $\ell_1$ norm not only provides a better estimate of $\hbX$, but it is also more resilient to denser signals $\bbX$.
The results also illustrate that, for the considered setup (10\% of errors in the graph), the alternatives are unable to properly identify the sources.

%\blue{[SRE: En este experimento el algoritmo ``Ye et al.'' no estima $\bbS$, por lo que entiendo que el error lo calculas con $\barbS$. Si es así indícalo en el experimento]}
\noindent \textbf{Test case 3}:
In the last test case, we analyze the performance of the graph denoising task when increasing the number of available graph signals $M$. Note that the approach in ``Ye et al.'' does not perform denoising, and therefore we set $\hbS = \barbS$ for this algorithm. We can see in Figure~\ref{F:exps}-(c) that, as expected, the normalized error of the GSO for the algorithms proposed in this work decreases as $M$ increases. The alternatives are again unable to properly denoise the GSO. It is worth mentioning that, even when only $M=20$ signals are observed (and hence $M=N$), our ``RBD-G-rew'' algorithm already obtains a normalized error of 0.1, providing a high-quality representation of the true GSO.
 
\section{Conclusion}\label{S:conc}

This work addressed the problem of blind deconvolution in the context of GSP when, instead of perfectly knowing the GSO, only an imperfect (perturbed) observation $\barbS$ is available. Our robust estimator was designed as the solution to an optimization problem where the sparse sources, the (inverse) filter, and the true graph were jointly learned. Two critical aspects of our design were: a) declaring the true graph as an optimization variable, and b) exploiting the fact that the inverse of the generative filter is a polynomial of the true graph via a commutativity bilinear constraint. These features resulted in an optimization fully formulated in the vertex domain and whose only source of non-convexity is the bilinear constraint, which in turn, is amenable to be solved via an alternating minimization approach. Preliminary numerical experiments showcased that our algorithm is able to identify the three variables in different scenarios. Future work includes the theoretical characterization of the proposed algorithm, designing more efficient optimization schemes to reduce the computational complexity, and using unrolling techniques to set the value of the regularization constants. 

\newpage

\bibliographystyle{IEEEbib}
\bibliography{myIEEEabrv,biblio}

\end{document}